\documentclass[11pt,amssymb,epsfig,amsmath]{article}

\usepackage[bookmarks=true]{hyperref}
\usepackage{graphicx,epsfig}
\usepackage{cite}
\usepackage[english]{babel}
\usepackage{amssymb,amsfonts,amsmath}

\newcommand{\be}{\begin{equation}}
\newcommand{\ee}{\end{equation}}
\newcommand{\ba}{\begin{eqnarray}}
\newcommand{\ea}{\end{eqnarray}}
\newcommand{\no}{\nonumber \\}
\newcommand{\gsim}{\mathrel{\hbox{\rlap{\lower.55ex \hbox {$\sim$}}
                   \kern-.3em \raise.4ex \hbox{$>$}}}}
\newcommand{\lsim}{\mathrel{\hbox{\rlap{\lower.55ex \hbox {$\sim$}}
                   \kern-.3em \raise.4ex \hbox{$<$}}}}

\def\del{{\partial}}

\def\roughly#1{\mathrel{\raise.3ex\hbox{$#1$\kern-.75em%
\lower1ex\hbox{$\sim$}}}}
\def\lsim{\roughly<}
\def\gsim{\roughly>}

\def\({\left(}
\def\){\right)}
\def\[{\left[}
\def\]{\right]}

\def\lsim{\mathrel{\rlap{\lower3pt\hbox{\hskip1pt$\sim$}}
     \raise1pt\hbox{$<$}}} 
\def\gsim{\mathrel{\rlap{\lower3pt\hbox{\hskip1pt$\sim$}}
     \raise1pt\hbox{$>$}}} 

\def\ka{\kappa}
\def\lam{\lambda}
\def\dlt{\delta}
\def\Dlt{\Delta}
\def\eps{\epsilon}
\def\omg{\omega}

\def\lv{\lvert}
\def\rv{\rvert}
\newcommand{\Gm}{\Gamma}
\newcommand{\lag}{\langle}
\newcommand{\rag}{\rangle}
\newcommand{\tb}{{\bar t}}
\newcommand{\Kb}{{\bar K}}
\newcommand{\deq}{\Relbar\hspace{-0.3cm}{}^d}
\newcommand{\pht}{{\tilde \phi}}

\begin{document}

\renewcommand{\thefootnote}{\arabic{footnote}}
\setcounter{footnote}{0}

\begin{flushright}
MPP-2011-134\\
TAUP-2938/11
\end{flushright}
\hfill {\today} \vskip 1cm

\begin{center}
{\LARGE\bf Time singularities of correlators from Dirichlet conditions in AdS/CFT}
\date{\today}

\vskip 1cm {
\large Johanna Erdmenger${}^a$\footnote{E-mail: jke@mpp.mpg.de},
Carlos Hoyos${}^b$\footnote{E-mail: choyos@post.tau.ac.il}\, and
Shu Lin${}^a$\footnote{E-mail: slin@mpp.mpg.de}\\
${}^a$Max-Planck-Institut f\"{u}r Physik (Werner-Heisenberg-Institut) \\ F\"{o}hringer Ring 6, 80805 M\"{u}nchen, Germany
\\
${}^b$
Raymond and Beverly Sackler School of Physics and Astronomy,\\  Tel-Aviv University,
Ramat-Aviv 69978, Israel 
 }


\end{center}




\vskip 0.5cm

\begin{abstract}

Within AdS/CFT, we establish a general procedure for obtaining the leading singularity
of two-point correlators involving operator insertions at different times.  The
procedure obtained is applied to operators dual to a scalar field which satisfies Dirichlet boundary conditions on an arbitrary time-like
surface in the bulk. We determine how the Dirichlet boundary conditions influence
the singularity structure of the field theory correlation functions.  
New singularities appear at boundary points connected by null
geodesics bouncing between the Dirichlet surface and the boundary. 
We propose that their appearance can be interpreted as due to a
non-local double trace deformation of the dual field theory, in which
the two insertions of the operator are separated in time. 
The procedure developed in this paper provides a technical tool
which may prove useful in view of describing holographic thermalization using gravitational collapse in AdS space.

\end{abstract}

\newpage

\renewcommand{\thefootnote}{\#\arabic{footnote}}
\setcounter{footnote}{0}



\section{Introduction}

One of the central questions in the physics of heavy ion collisions is
the mechanism of thermalization. A short thermalization time of the
order of $0.5fm/c$ is suggested by successful hydrodynamical
descriptions of elliptic flow data, yet its mechanism remains poorly
understood. Recently, there have been increasing amount of efforts in
the studies of thermalization with both phenomenological and theoretical
approaches. Different scenarios have been proposed to describe the
thermalization process. We will mention a few of them here: In
\cite{DGV}, the role of quantum fluctuations were emphasized in the
equilibration of initial coherent fields in a Color-Glass-Condensate
approach. In \cite{moore}, the mechanism driving the system to
equilibrium was attributed to an interplay between the plasma
instability and Bjorken expansion. A possible Bose-Einstein
condensation of gluons 
at the onset of equilibration has also been proposed \cite{liao}.

The abovementioned studies are all based on a weak coupling
picture. Studies of thermalization in the strong coupling regime
have been mainly carried out in the framework of the AdS/CFT
correspondence. 
These include gravitational collapse models
\cite{uppsala,giddings,LS,Minwalla,chesler,zayas}  as well as
gravitational shock wave collision models
\cite{GPY,grazing,GPY2,KL,mozo,bagrov,LScc,taliotis}. While these works have been  successful in
describing the formation of a black hole in AdS, which corresponds to
the equilibration of gauge fields, the evolution of correlation
functions in the process of thermalization remains largely
unexplored. Recently, there has been some initial progress in this
direction \cite{LS,kovchegov,headrick,teaney,11authors}. In
particular, \cite{11authors} studied various quantities, including
equal-time correlators, Wilson loops and entanglement entropy in a
thermalization process of a strongly coupled gauge theory within a
gravitational collapse model.
Moreover, in \cite{11authors} a ``top-down'' scenario was proposed
that gives rise to
a thermalization cascade from UV to IR modes
for the dual strongly coupled gauge theory.
Whether this cascade is a universal feature of strongly coupled
theories with a holographic dual remains to be seen. 
This behaviour is in contrast  to the ``bottom-up'' scenario established in weak
coupling analysis \cite{baier}, where the IR mode equilibrates first,
followed by a loss of energy of the UV mode.

While the observables that have been studied previously contain
valuable information about the state of the system at a given time,
more information can be extracted from further observables that are
not evaluated at fixed times, for instance from correlators involving
operators at different times. In contrast to systems at thermal
equilibrium, the response of the system is not determined by the
fluctuation-dissipation theorem. Moreover, correlation of operators at
different times depends not only on the time difference but also on
the initial time. This kind of observables are
particularly valuable when studying the long-time behaviour of the system following an initial perturbation.

Interestingly, the thermalization mechanism has also been investigated
recently in the condensed matter community, mainly due to the
advancement of experimental tools for the study of ultracold atoms and
quantum phase transitions \cite{CC_ref}. A canonical question in
condensed matter physics is how  a system behaves under a quantum
quench. This refers to the behavior of a system after a sudden change
of a parameter in the original Hamiltonian, after which it undergoes a
unitary evolution according to the new Hamiltonian. This can lead to
thermalization.  General expressions for one-point and two-point
correlators after a quantum quench have been obtained for $d=1$ space
dimensions using two-dimensional
CFT techniques in the seminal work \cite{CC}. Moreover, these authors
argue that the two-point correlator has the same generic behavior also
in higher dimensions. These novel results have triggered an extensive
study  of quantum quench also in higher-dimensional CFT by methods of the
AdS/CFT correspondence, with many interesting
results\cite{lopez,johnson,das,vakkuri}. However, the mechanism of 
thermalization in higher dimension is still unclear, as opposed to its
$(1+1)$-dimensional  counterpart.

In view of these results, in this paper we establish a new technical
tool which we expect to be useful in further studies of
thermalization and out-of equilibrium processes in general. 
We focus on a particular class of observables, i.e. the spatially
integrated unequal-time correlator  $\lag\int
d^{d-1}xO(t,x)O(t',0)\rag$. We study the singularity structure of the
correlator in the framework of AdS/CFT.  To be specific, we consider
the operator dual to a bulk scalar. Extensions to other operators are
straightforward. 

A standard thermalization model in AdS/CFT consists of a collapsing
shell of matter in the bulk that eventually forms a black hole. In
order to illustrate our method, here however we consider a simpler
setup where the shell of matter has a completely reflective surface
with Dirichlet boundary conditions -
a mirror - and does not backreact on the geometry. We will not make
assumptions about the evolution of the collapse, but we work out
simple examples explicitly. Although this is a crude approximation to
the real problem, it may capture some of the features of a real
collapsing shell before the formation of a horizon starts. In
particular, we obtain useful results on divergence matching which are
expected to generalize to boundary conditions other than Dirichlet. We
plan to apply these result to the collapsing shell geometry with
appropriate boundary conditions in the future.

This work is based on the previous work \cite{hoyos}
by Amado and one of the present authors and \cite{ELN} by Ngo and the
other two of the present authors. In \cite{hoyos}, the scalar wave in
AdS space with a static mirror, which provides a Dirichlet boundary
condition, has been considered. The corresponding two-point correlator on
the boundary field theory showed singularities when the insertion
times of the operators are connected a bouncing null geodesics between
the mirror and the boundary. In \cite{ELN}, a mirror trajectory which
breaks time translational invariance but preserves scaling invariance
has been considered. An explicit evaluation of the two-point
correlator in this moving mirror setting  confirmed that the structure
of the singularities in the correlator is consistent with a bouncing
null geodesic picture. These two examples are realizations of the
bulk-cone singularity conjecture  \cite{HLR} in Poincar\'e
coordinates. According to this conjecture, singularities in boundary
correlators appear whenever two insertion points at the boundary are
connected by null geodesics in the bulk. This is based on the
observation that bulk correlators are singular on null surfaces and
those are inherited by the boundary correlators. Such singularities
can appear inside the boundary lightcone, so they contain both
information about the bulk that can be used to learn about its causal
structure, and dynamical information about the boundary theory. This
was applied to study the stages prior to horizon formation by a
collapsing shell in the original work proposing the bulk-cone
singularity conjecture \cite{HLR}.

In previous works the bulk-cone singularity was used to determine the location of time singularities of dual correlators. In the present work, we extend the analysis to determine the precise form of the singularities and their coefficients in the presence of an arbitrary moving mirror in AdS.

We expect that these results will provide an essential tool for
the future study of the thermalization process in the gravitational
collapse model. Knowledge of the singularity structure 
of correlators as described above will provide useful information
about the behavior of strongly coupled gauge theories far away from
equilibrium, in particular as far as decoherence is concerned.

This paper is organized as follows: In section 2, we present and
sketch the derivation of the divergence matching method. We find a set
of recursion equations relating the most singular part of the
two-point correlator at adjacent singularities. The initial conditions
to the recursion equations are obtained from the vacuum bulk-boundary
propagator. Solving the recursion equations with the initial
conditions allows us to determine the precise form for all
singularities. In sections 3 and 4, we test of the divergence matching
method with explicit evaluations of the two-point correlator in the
cases of static mirror \cite{hoyos} and mirror in constant motion \cite{ELN}. In section 5, we interpret the Dirichlet boundary condition as non-local double trace deformation in the dual field theory and argue that the non-local double trace deformation generically leads to the emergence of new singularities in the two-point correlator. In section 6, the form of the singularities is explained in terms of spectral decomposition for the cases of the static mirror and the scaling mirror. We end with conclusion and outlook in section 7.

\section{Singularities for arbitrary mirror trajectories}

Our goal is to find the coefficient of singularities of the two-point correlator of a scalar operator, using holography and imposing a Dirichlet boundary condition on an arbitrary (time-like) surface $z=f(t)>0$, for the dual scalar field in an $AdS$ geometry
$$
ds^2=\frac{1}{z^2}\left(dz^2+\eta_{\mu\nu}dx^\mu dx^\nu \right).
$$
In general, to solve the equations of motion with an arbitrary
condition of this kind is too difficult, and only in some simple cases
explicit solutions are known. However, in order to find the
singularities it is not necessary to know the full solution, the
leading terms in a WKB approximation are sufficient. As it has been shown in several examples, the localization of singularities can be obtained from null geodesics bouncing on the mirror \cite{hoyos, ELN}.

Here, in order to study the effect of the Dirichlet boundary condition, we introduce by hand a potential barrier localized at the position of the mirror in the equations of motion of the scalar field, and then we take the strength to infinity. For simplicity, we will focus on a massless scalar field. 
The equations of motion in the presence of the potential are
\begin{equation}\label{KG_eq}
\frac{1}{\sqrt{-g}}\partial_\mu\left( \sqrt{-g} g^{\mu\nu}\partial_\nu\right)\Phi +V\Phi=0,
\end{equation}
where $V=V_0\delta(z-f(t))$. 
\eqref{KG_eq} is a Klein-Gordon equation in the presence of a potential. Experience from Quantum Mechanics tells us that the limit $V_0\to \infty$ corresponds to the Dirichlet boundary condition at $z=f(t)$.\footnote{Examples can be found in Appendices \ref{apC} and \ref{apB}.} 

The computation of the two-point correlator is equivalent to finding a bulk-boundary 
correlator, which satisfies the  condition:
\ba\label{eqV}
(\square+V)G(t,z,t')=0,
\text{with}\, G(t,z\rightarrow0,t')\rightarrow \dlt(t-t').
\ea
We restrict ourselves to spatially homogeneous solutions to  \eqref{eqV}, which will
lead to the spatially integrated two-point correlator \cite{ELN}. With this simplification,
the Laplacian operator in $AdS_{d+1}$ reduces to 
$\square=-z^2\del_t^2+z^2\del_z^2+z(1-d)\del_z$. 
 \eqref{eqV} can be reformulated as the integral equation
\ba\label{master}
G(t,z,t')=G_0(t,z,t')-\int G_{bb}(t,z,t'',z'')V(t'',z'')\sqrt{-g(z'')}G(t'',z'',t')dt''dz'',
\ea
where $G_0(t,z,t')$ and $G_{bb}(t,z,t'',z'')$ are the bulk-boundary propagator and 
the bulk-bulk propagator in the absence of a potential. They are defined as follows:
\ba
&&\square G_0(t,z\rightarrow0,t')=0 \no
&&\text{with}\;G_0(t,z\rightarrow0,t')\rightarrow \dlt(t-t')\;, \\
&&\square G_{bb}(t,z,t',z')=\frac{1}{\sqrt{-g}}\dlt(t-t'')\dlt(z-z'') \no
&&\text{with}\;G_{bb}(t,z\rightarrow0,t',z')\rightarrow 0.
\ea
Let us focus on the time-ordered propagators. The explicit expressions are given by \cite{tasi}:
\ba\label{eq:noVcorr}
&&G_0(t,z,t')=\frac{i}{\pi}\frac{\Gm(\frac{d+1}{2})\Gm(\frac{1}{2})}{\Gm(\frac{d}{2})}\frac{z^d}{(-(t-t')^2+z^2+i\eps)^{\frac{d+1}{2}}}\;, \\
&&G_{bb}(t,z,t',z')=-\frac{i}{2\pi}(zz')^{\frac{d-1}{2}}Q_{\frac{d-1}{2}}\left(\frac{z^2+z'{}^2-(t-t')^2+i\eps}{2zz'}\right).
\ea
Note the $i\eps$ prescription is chosen for time ordered correlators (Feynman). For $t>t'$, $G_0(t,z,t')$ contains only contribution from positive frequency modes, and for $t<t'$, $G_0(t,z,t')$ contains only contribution from negative frequency modes. Similarly, for $t>t'$, $G_{bb}(t,z,t',z')$ is a propagator for positive frequency modes, and for $t<t'$, $G_{bb}(t,z,t',z')$ is a propagator for negative frequency modes. These properties will be crucial in the analysis below. The delta function in the potential $V$ forces the integration in \eqref{master} to be performed along the mirror trajectory.

\begin{figure}
\includegraphics[width=0.45\textwidth]{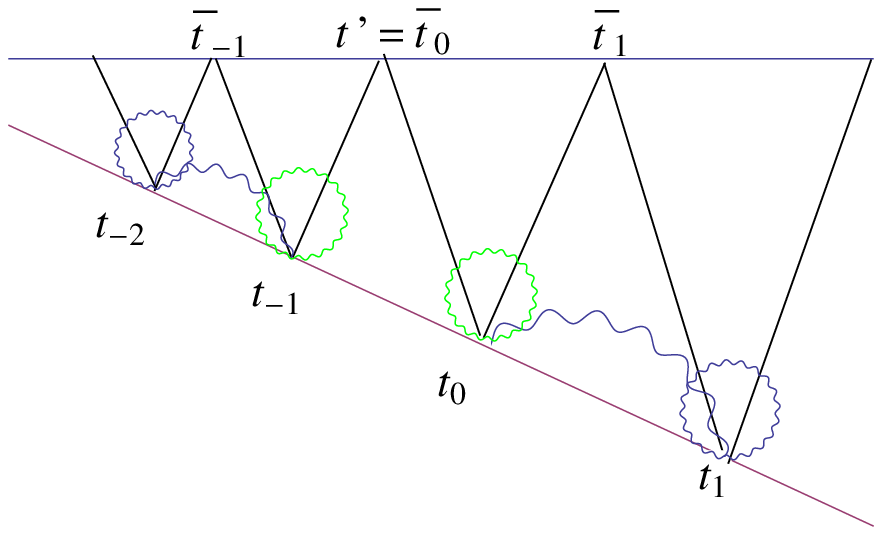}
\includegraphics[width=0.45\textwidth]{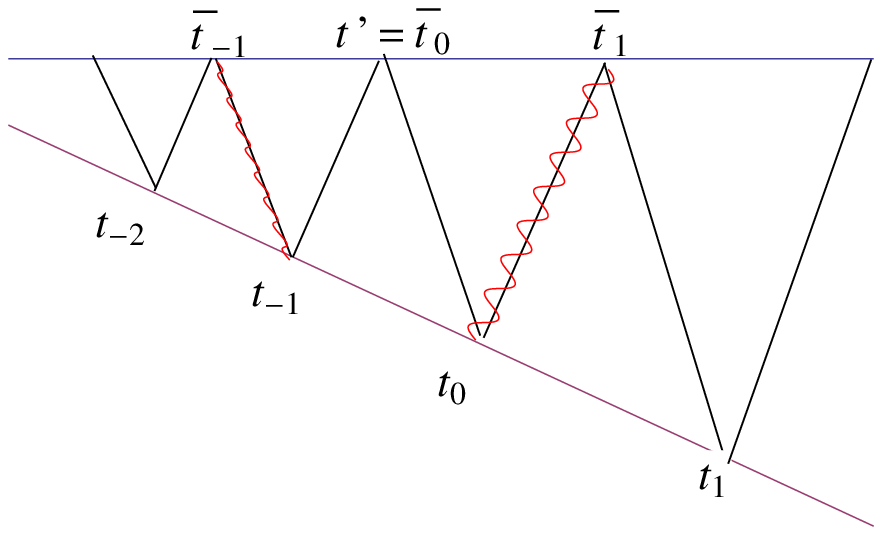}
\caption{\label{wavy}(color online) The figure on the left illustrates the
  cancellation of the most singular parts. The wavy lines denote the
  $K(t\to t_m,t''\to t_n)$. For $m>0$, the contributions from the blue
  wavy loop and the blue arc on the right cancel against each other,
  corresponding to $n=m$ and $n=m-1$. For $m>0$, the contributions
  from the blue wavy loop and the blue arc on the left cancel,
  corresponding to $n=m$ and $n=m+1$. For $m=0,-1$, only $n=m$
  contributes. These contributions are depicted as green wavy loops, and 
cancel the most singular part of $G_0(t,z,t')$. The figure on the right
  illustrates how the most singular parts are propagated to the
  boundary: For $m>0(m<0)$, only $n=m-1(n=m)$ contributes to the
  two-point correlator, as indicated by the red wavy line on the
  right(left).  }
\end{figure}

Notice the following: $G_0$ and $G_{bb}$ are independent of $V_0$, so
in order for \eqref{master} to be consistent in the $V_0\to \infty$
limit, two conditions must be satisfied, the first is that $G\sim
1/V_0$ and the second is that the $O(V_0^0)$ contributions from $G_0$
and the integral term cancel out.\footnote{Then, in this limit we
  should normalize the correlator by multiplying it by a factor of $V_0$ in order to obtain a finite result.} 
From the last condition one can obtain the coefficients of singularities. The idea works as follows: Let us define $G=\frac{1}{V_0}G_{LO}+O(V_0^{-2})$, then
\begin{align}\label{K_def}
&\int G_{bb}(t,z,t'',z'')V(t'',z'')G(t'',z'',t')\sqrt{-g(z'')}dt''dz'' \no
=&\int G_{bb}(t,z,t'',f(t'')) G_{LO}(t'',f(t''),t')\sqrt{-g(f(t''))}dt''+O(V_0^{-1}).
\end{align}
Then, to the order $O(V_0^0)$, we have the following conditions
\begin{align}
\label{recursion3}
&G_0(t,z,t')-\int G_{bb}(t,z,t'',f(t''))G_{LO}(t'',f(t''),t')\sqrt{-g(f(t''))} dt''= 0.
\end{align}
We first look at  \eqref{K_def} along the trajectory of the mirror: $z=f(t)$, Generically, we expect $G_{LO}(t'',f(t''),t')$ to be singular whenever $t''$ is connected by a null geodesic to $t'$, $t''\to t_n$, as remarked in Fig.\ref{wavy}. Furthermore, $G_{bb}$ has a singularity whenever the points $(t,f(t))$ and $(t'',f(t''))$ are connected by a null geodesic bouncing once at the boundary. One can see this from \eqref{eq:noVcorr}, the function $Q_{\frac{d-1}{2}}(x)$ has logarithmic singularities at $x=\pm 1$. This gives the condition
\begin{equation}
f(t)^2+f(t'')^2-(t-t'')^2=\pm 2 f(t) f(t'') \ \Rightarrow \ (f(t)\mp f(t''))^2=(t-t'')^2.
\end{equation}
Singularities appear at $t=t''$ and $t=t''\pm \Delta t$, where $\Delta t=f(t)+f(t'')$ is the time it takes a null ray to go from $z=f(t)$ to $z''=f(t'')$ bouncing once at the boundary. e.g., when \eqref{K_def} is evaluated at $(t_0,f(t_0))$ the term in the integral has a singular contribution from $(t_{-1},f(t_{-1})$, $(t_0,f(t_0))$ and from $(t_1,f(t_1))$, the next point in the mirror connected by a null geodesic bouncing at the boundary. The convolution of $G_{bb}(t,f(t),t'',f(t''))$ and $G_{LO}(t'',f(t''),t')$ close to $t''\to t_n$ gives the most singular part, which is to be cancelled by the most singular part of $G_0(t,f(t),t')$. Note that $G_0$ is singular only when $t\to t_0,t_{-1}$. We arrive at the following matching conditions:
\begin{subequations}
\begin{align}
&\sum_{n=m-1}^{m+1}\int^{t_n}  K(t\to t_m,t'')G_{LO}(t'',t') dt''\deq\, 0, \quad m\neq 0,-1, \label{m_rest}\\
&G_0(t\to t_0,f(t),t')-\sum_{n=-1}^{1}\int^{t_n}  K(t\to t_0,t'')G_{LO}(t'',t') dt''\deq\, 0, \quad m=0, \label{m_0}\\
&G_0(t\to t_{-1},f(t),t')-\sum_{n=-2}^{0}\int^{t_n}  K(t\to t_{-1},t'')G_{LO}(t'',t') dt''\deq\, 0, \quad m=-1, \label{m_m1}
\end{align}
\end{subequations}
where $K(t,t'')=G_{bb}(t,f(t),t'',f(t''))\sqrt{-g(f(t''))}$. The symbol $\deq$ means the equality holds as far as the most singular part is concerned. The superscript of the integration sign means that the main contribution to the integral comes from the singular behavior close to the singularities $t_n$. Focusing
on the most singular part allows us to identify a set of discrete conditions, thus significantly simplifying the problem. We will illustrate how this works with the explicit examples of a static mirror \cite{hoyos} and a mirror with scaling trajectory \cite{ELN}.

Schematically, the singularities in $G_0(t,f(t),t')$ get propagated through \eqref{m_rest}-\eqref{m_m1} to $G_{LO}(t'',f(t''),t')$. Since we focus on the time-ordered the propagator, $G_{LO}$ contains only positive (negative) frequency modes for $t''>t'$($t''<t'$). Therefore, the most singular part of $G_{LO}$ assumes a similar form as $G_0$:
\ba\label{Glo}
G_{LO}(t,t')=G_{LO}(t=t_m(1+y),z=f(t),t')\deq 
\left\{\begin{array}{l@{\quad\quad} l}
\frac{g_m}{(-iy+\eps)^{c+1}} &m\le -1\\
\frac{g_m}{(iy+\eps)^{c+1}} &m\ge 0
\end{array}
\right.,
\ea
where $c=\frac{d+1}{2}$ and $g_m$ are some constants to be fixed by the matching procedure.

Now we wish to determine $g_m$ through the recursion relations 
(\ref{m_rest})-(\ref{m_m1}). 
Close to the points where null geodesics bounce on the mirror $t=t_m(1+x)$, $z\simeq f(t_m)+t_m f'(t_m)x$ and $t''= t_n(1+y)$, $z''\simeq f(t_n)+t_n f'(t_n)y$, with $x,y\ll 1$, the most singular part of $K(t,t'')$ is given by
\begin{align}\label{K_singular}
K(t=t_m(1+x)),t''=t_n(1+y))&\deq -\frac{i}{2\pi}\frac{z_m}{z_n}^{\frac{d-1}{2}}\frac{1}{z_n^2} \times \no
&\left\{\begin{array}{l@{\quad\quad} l}
A_4e^{\frac{-i\pi(d+1)}{2}}\ln\Big(-\frac{1+f'(t_m)}{1-f'(t_n)}\frac{t_m}{t_n}x+y-i\eps\Big)& n=m+1\\
A_4e^{\frac{-i\pi(d+1)}{2}}\ln\Big(\frac{1-f'(t_m)}{1+f'(t_n)}\frac{t_m}{t_n}x-y-i\eps\Big)& n=m-1\\
A_4\ln\Big(-(x-y)^2+i\eps\Big)& n=m
\end{array}
\right.  ,
\end{align}
where we have defined $z_m=f(t_m)$, $z_n=f(t_n)$ and $A_4=-\frac{\Gm(\frac{d+1}{2})\Gm(\frac{1}{2})}{2^{\frac{d+1}{2}}\Gm(\frac{d+1}{4})\Gm(\frac{d+3}{4})}$.
Inserting (\ref{Glo}) and (\ref{K_singular}) into (\ref{m_rest}), and using the integrals in Appendix \ref{apA}, we find that for $m>0$, only $n=m,m-1$ contribute, while for $m<-1$, only $n=m,m+1$ contribute.
The physical reason for this is that $K(t\to t_m,t''\to t_n)$ at
$n=m-1$ and $n=m+1$ are  propagators for positive or negative
frequency modes only. The convolution of a positive (negative)
propagator with a singularity due to negative (positive) frequency
modes always gives a vanishing result. On the other hand, $K(t\to
t_m,t''\to t_n)$ at $n=m$ is a propagator for both positive and
negative frequency modes, thus always contributes to the
integral. This mechanism is precisely encoded in the vanishing of the
integrals in Appendix \ref{apA}. We have indicated the cancellation of
the nonvanishing contributions in the left panel of Fig.\ref{wavy}. As
a result, we can derive the following relations among the $g_m$
defined in \eqref{Glo}:
\ba\label{gm}
&e^{i\pi c}(1-f'(t_m))^c\(\frac{t_m}{z_m}\)^{c+1}g_m+(1+f'(t_{m-1}))^c\(\frac{t_{m-1}}{z_{m-1}}\)^{c+1}g_{m-1}=0, &m> 0 ,\no
&(1-f'(t_m))^c\(\frac{t_m}{z_m}\)^{c+1}g_m+e^{i\pi c}(1+f'(t_{m-1}))^c\(\frac{t_{m-1}}{z_{m-1}}\)^{c+1}g_{m-1}=0, &m\leq -1,
\ea
The initial condition for (\ref{gm}) follows from (\ref{recursion3}).
Note that the singular behavior of $G_0$ is given by
\begin{align}\label{G0_singular}
G_0(t=t_m(1+x),t')\deq\frac{i}{\pi}\frac{\Gm(\frac{d+1}{2})\Gm(\frac{1}{2})}{\Gm(\frac{d}{2})}\frac{z_m^d}{(2t_mz_m)^c}
\left\{\begin{array}{l@{\quad\quad} l}
(1-f'(t_m))^{-c}(-x+i\eps)^{-c}& m=0\\
(1+f'(t_m))^{-c}(x+i\eps)^{-c}& m=-1
\end{array}
\right. .
\end{align}
Using (\ref{Glo}), (\ref{K_singular}) and (\ref{G0_singular}), as well
as the integrals in Appendix \ref{apA}, we find that only $n=m$
contributes to the integral. The physical reason is again that the
singularities due to positive and negative frequency modes cannot meet
the correct propagators for the cases $n=m\pm1$. As a result, we can easily obtain $g_0$ and $g_{-1}$ as follows:
\begin{align}\label{g01}
&g_0=\frac{\Gm(\frac{d+1}{2})\Gm(\frac{1}{2})}{\pi\Gm(\frac{d}{2})}\frac{c}{(2i)^c A_4}\left(\frac{z_0}{t_0}\right)^{c+1}(1-f'(t_0))^{-c} \\
&g_{-1}=\frac{\Gm(\frac{d+1}{2})\Gm(\frac{1}{2})}{\pi\Gm(\frac{d}{2})}\frac{c}{(2 i)^c A_4}\left(\frac{z_{-1}}{t_{-1}}\right)^{c+1}(1+f'(t_{-1}))^{-c},
\end{align}
where we have used $i=e^{\frac{i\pi}{2}}$ to simplify the notation.
To obtain the two point-correlator in the gauge theory, we need to propagate
singularities at $t=t_n$ to the boundary, where they will appear at different times $t=\tb_n$. To this end, we need to look at \eqref{master} in the limit $z\to 0$. Note that the time-ordered correlator $\lag T O(t)O(t')\rag=\lim_{z\to 0}\frac{G(t,z\to 0,t')}{z^d}$. The most singular part of the two-point correlator is given by:
\begin{align}\label{master_matching}
\lag TO(t)O(t')\rag\vert_{t\to \tb_n}\deq-\sum_{n=m-1}^m\int^{t_n}\Kb(t\to\tb_n,t'')G_{LO}(t'',t')dt'',
\end{align}
where
\ba
\Kb(t,t'')=\lim_{z\to 0}\frac{G_{bb}(t,z,t'',f(t''))}{z^d}\sqrt{-g(f(t''))}.
\ea
It is not difficult to work out the
most singular part of $\Kb(t,t'')$ for an arbitrary mirror trajectory:
\begin{align}\label{Kb_singular}
\Kb(t=\tb_m(1+x),t''=t_n(1+y))&\deq -\frac{i}{\sqrt{\pi} (2 z_n)^{c+1}}\frac{\Gm(\frac{d+1}{2})}{\Gm(\frac{d+2}{2})} \times \no
&\left\{\begin{array}{l@{\quad\quad} l}
\(\tb_m x-(1-f'(t_n)) t_n y+i\eps\)^{-c}& n=m\\
\(-\tb_mx+(1+f'(t_{n}))t_n y+i\eps\)^{-c}& n=m-1
\end{array}
\right. .
\end{align}
Finally, inserting \eqref{Glo} and \eqref{Kb_singular} into \eqref{master_matching} and using the integrals in Appendix \ref{apA}, we obtain the most singular part of the two-point correlators. A similar mechanism is at work again: Time-ordered correlators only propagate singularities due to either positive  or negative frequency modes, thus only one out of two contributions survives, which is indicated in the right panel of Fig. \ref{wavy}. We collect the final results for the most singular part of the two-point correlator in the following:
\begin{align}\label{correlator_f}
&\langle TO(t=\tb_m(1+x))O(t')\rangle \, \deq  \,\frac{\alpha_{m-1} g_{m-1}}{(-x+i\eps)^{2c}}, \quad m>0,\no
&\langle TO(t=\tb_m(1+x))O(t')\rangle\, \deq \, \frac{\beta_m g_m}{(x+i\eps)^{2c}}, \quad m<0,
\end{align}
where
\begin{align}\label{correlator_coef}
&\alpha_{m-1}=-\frac{i^{c+1}}{\sqrt{\pi}}\sin(\pi c)  B(2c,1-c) \frac{\Gm(\frac{d+1}{2})}{\Gm(\frac{d+2}{2})} \(\frac{t_{m-1}}{z_{m-1}}\)^{c+1}\frac{(1+f'(t_{m-1}))^c}{(2\tb_m^2)^c}, \quad m>0,\\
&\beta_m=-\frac{i^{c+1}}{\sqrt{\pi}}\sin(\pi c) B(2c,1-c)\frac{\Gm(\frac{d+1}{2})}{\Gm(\frac{d+2}{2})} \(\frac{t_m}{z_m}\)^{c+1}\frac{(1-f'(t_m))^c}{(2\tb_m^2)^c}, \quad\quad\quad m<0.
\end{align}

Although these expressions look  complicated, we can easily extract interesting physics by comparing the ratio between the residues of two consecutive singularities. To this end, we normalize the two-point correlators \eqref{correlator_f} as:
\begin{align}\label{correlator_n}
&\langle TO(t=\tb_m(1+x))O(t')\rangle\deq \frac{\alpha_{m-1} g_{m-1}\tb_m^{2c}}{(-t+\tb_m+i\eps)^{2c}} ,\quad m>0,\no
&\langle TO(t=\tb_m(1+x))O(t')\rangle\deq \frac{\beta_m g_m\tb_m^{2c}}{(t-\tb_m+i\eps)^{2c}}, \quad m<0.
\end{align}
The ratio of consecutive residues are given by:
\begin{equation}\label{residues}
R_m\equiv \left\{ \begin{array}{lc} \frac{\alpha_{m} g_{m}\tb_{m+1}^{2c}}{\alpha_{m-1} g_{m-1}\tb_m^{2c}}=-e^{-i\pi c} \left(\frac{1+f'(t_m)}{1-f'(t_m)}\right)^c,  & m>0, \\
\frac{\beta_m g_m\tb_m^{2c}}{\beta_{m-1} g_{m-1}\tb_{m-1}^{2c}}=-e^{-i\pi c} \left(\frac{1+f'(t_{m-1})}{1-f'(t_{m-1})}\right)^c ,& m<0
\end{array}\right..
\end{equation} 
If one introduces a signal at the boundary it will have replicas at the time $\tb_m$. The strength of the replicas depends on the relative value of the residues at future singularities: It becomes stronger when $|R_m|>1$ and weaker when $|R_m|<1$.

\section{Simple example: static mirror}

Let us see how the procedure works in the case of a static
mirror, as  first studied in \cite{hoyos}. The mirror sits at a constant radial 
coordinate: $f(t)=z_s$. It is straightforward to determine the singular points
from geometric optics:
\begin{align}
&t_m=t'+(2m+1)z_s,\,z_m=z_s\\
&\tb_m=t'+2mz_s.
\end{align}
Since $f'(t)=0$, \eqref{residues} predicts that the signal is always reflected with the same amplitude for a static mirror:
\begin{equation}
|R_m|\equiv 1, \ \ m>0.
\end{equation}

Note that along the mirror trajectory $G(t,z=f(t),t')\sim\frac{1}{V_0}$. The left hand side of \eqref{master} is $\frac{1}{V_0}$ suppressed as compared to the right hand side (rhs) in the limit $V_0\to\infty$, which allows us to focus on the rhs to leading order (LO) in $V_0$. Along the mirror trajectory, $G(t,z=f(t),t')$ is most singular when $t\to t_m$ and 
$K(t,z,t'')$ is logarithmically divergent when $t$ and $t''$ are
adjacent points on the mirror trajectory  hit by the light ray,
i.e. $t''\to t_n(n=m-1,m,m+1)$, or when $t$ and $t''$ coincide. As a result, (\ref{master}) can be written as:
\begin{align}\label{master_GK}
G_0(t\to t_m,z,t')-\sum_{n=m-1}^{m+1}\int^{t_n} K(t\to
t_m,z,t'')G_{LO}(t'',t')dt'' \, \deq \, 0.
\end{align}

To proceed on, let us assume the most singular part of $G_{LO}(t,z,t')$ takes the following form along the mirror trajectory:
\ba\label{G_singular}
G_{LO}(t=t_m(1+y),t')\deq 
\left\{\begin{array}{l@{\quad\quad} l}
\frac{g_m}{(-iy+\eps)^{c+1}}, &m\le -1,\\
\frac{g_m}{(iy+\eps)^{c+1}}, &m\ge 0.
\end{array}
\right.
\ea
The difference in $i\eps$ prescription indicates that the
singularities are due to positive and negative frequency modes,
respectively. As noted in the previous section, singularities from
positive frequency contribute to the integral only when $n=m,m-1$,
while singularities from negative frequency contribute to the integral
only when $n=m,m+1$. We are now ready to insert the singular forms of $G(t\to t_m,t')$ and $K(t\to t_m,t''\to t_n)$ from (\ref{G_singular}) and (\ref{K_singular}) into (\ref{master_GK}). We obtain the recursion relations among $g_m$ in (\ref{gm}) as well as the initial conditions $g_0$ and $g_{-1}$ in (\ref{g01}).
For the case of a static mirror, we can easily solve for the
coefficients and obtain
\begin{align}\label{gm_s}
&g_m=\(\frac{-1}{e^{i\pi c}}\)^m\(\frac{t_0}{t_m}\)^{c+1}g_0,\quad\quad\quad m>0,\no
&g_m=\(\frac{-1}{e^{i\pi c}}\)^{-1-m}\(\frac{t_{-1}}{t_m}\)^{c+1}g_{-1},\quad m<-1,
\end{align}
with the explicit initial conditions
\begin{align}\label{g01_s}
&g_0=\frac{\Gm(\frac{d+1}{2})\Gm(\frac{1}{2})}{\pi\Gm(\frac{d}{2})}\frac{c}{A_4}\frac{z_s^{c+1}}{(2i)^ct_0^{c+1}}\,
,\no
&g_{-1}=\frac{\Gm(\frac{d+1}{2})\Gm(\frac{1}{2})}{\pi\Gm(\frac{d}{2})}\frac{c}{A_4}\frac{z_s^{c+1}}{(2i)^ct_{-1}^{c+1}}.
\end{align}

To calculate the two-point correlator, we take the $z\to 0$ limit in \eqref{master}, close to the boundary. Note the two-point correlator is simply
\ba
\lag TO(t)O(t')\rag=\lim_{z\to0}\frac{G(t,z,t')}{z^d}.
\ea
The most singular part of the two-point correlator can be readily obtained from (\ref{master}),
\ba\label{Too}
\lag TO(t)O(t')\rag\deq-\int\Kb(t,t'')G(t'',t')dt''.
\ea
We have dropped the $G_0$ term since it is the vacuum piece of the correlator, which only has a trivial lightcone singularity at $t=t'$.
With the most singular part of $G_{LO}(t\to t_m,t'')$ obtained above and noting that $\Kb(t,t'')$ is most singular when $t$ and $t''$ are connected by light ray (null geodesics), we can write (\ref{Too}) as
\ba\label{Too2}
\lag TO(t\to\tb_m)O(t')\rag\deq-\sum_{n=m-1}^m\int^{t_n}\Kb(t\to\tb_m,t'')G(t'',t')dt''\quad m\ne 0.
\ea
Using the singular forms of $G_{LO}(t''\to t_n,t')$ and $\Kb(t\to\tb_m,t''\to t_n)$ and taking into account that only one out of the two terms contributes, we readily reduce \eqref{Too2} to \eqref{correlator_f}.
Inserting \eqref{gm_s} and \eqref{g01_s} into \eqref{correlator_f}, we finally obtain the most singular part of the correlator for the state defined by the static mirror:
\begin{align}\label{correlator_s}
&\langle TO(t=\tb_m(1+x))O(t')\rangle\deq-\frac{i}{2\pi}\frac{\sqrt{\pi}\Gm(\frac{d+1}{2})}{\Gm(\frac{d+2}{2})}\frac{1}{(2\tb_m^2)^c}\(e^{-i\pi c}-e^{i\pi c}\) \no
&\times i^{c+1}B(2c,1-c)\(\frac{-1}{e^{i\pi
    c}}\)^{m-1}\frac{\Gm(\frac{d+1}{2})\Gm(\frac{1}{2})}{\pi\Gm(\frac{d}{2})}\frac{c}{A_4}\frac{1}{(2i)^c(-x+i\eps)^{2c}}
\, ,\quad m>0,\no
&\langle TO(t=\tb_m(1+x))O(t')\rangle\deq-\frac{i}{2\pi}\frac{\sqrt{\pi}\Gm(\frac{d+1}{2})}{\Gm(\frac{d+2}{2})}\frac{1}{(2\tb_m^2)^c}\(e^{-i\pi c}-e^{i\pi c}\) \no
&\times i^{c+1}B(2c,1-c)\(\frac{-1}{e^{i\pi
    c}}\)^{-m-1}\frac{\Gm(\frac{d+1}{2})\Gm(\frac{1}{2})}{\pi\Gm(\frac{d}{2})}\frac{c}{A_4}\frac{1}{(2i)^c(x+i\eps)^{2c}}
\, ,\quad m<0.
\end{align}
We can cross-check the above results by direct evaluation of $G_{LO}$ and the two-point correlator $\lag TO(t)O(t')\rag$. We have done this in Appendix \ref{apC} and find perfect agreement!

\section{Non-trivial example: mirror along scaling trajectory}

In this section, we will test our procedure with a non-trivial example. A non-trivial yet still analytically tractable example was studied in \cite{ELN}. It corresponds to a mirror with scaling trajectory $z=\frac{t}{u_0}$ with $u_0>1$. The trajectory breaks the translational symmetry in time, but preserves the scaling symmetry.

As for the static mirror, the location of singular points can deduced from geometric optics:
\begin{align}
&t_m=t'\frac{u_0}{u_0-1}\left(\frac{u_0+1}{u_0-1}\right)^m,\\
&\tb_m=t'\left(\frac{u_0+1}{u_0-1}\right)^m.
\end{align}
Using \eqref{residues} we find the ratio of consecutive residues is given by:
\begin{equation}
|R_m|=\left(\frac{u_0+1}{u_0-1}\right)^{c}.
\end{equation}
So the residues grow when the mirror moves away from the boundary. We will comment more on this in Section \ref{sec}.

Introducing now the expression for the trajectory $f(t)=\frac{t}{u_0}$ in (\ref{m_rest})-(\ref{m_m1}), we obtain the recursion equations
\begin{align}\label{scaling_gm}
&e^{i\pi c}(u_0-1)^cg_m+(u_0+1)^cg_{m-1}=0\quad m>0 \\
&e^{i\pi c}(u_0+1)^cg_m+(u_0-1)^cg_{m+1}=0\quad m<-1,
\end{align}
together with the initial conditions
\begin{align}
&g_0=\frac{\Gm(\frac{d+1}{2})\Gm(\frac{1}{2})}{\pi\Gm(\frac{d}{2})}\frac{c}{A_4}\frac{1}{(2i)^cu_0(u_0-1)^c} \\
&g_{-1}=\frac{\Gm(\frac{d+1}{2})\Gm(\frac{1}{2})}{\pi\Gm(\frac{d}{2})}\frac{c}{A_4}\frac{1}{(2i)^cu_0(u_0+1)^c} .
\end{align}
Solving the recursion equations, we obtain
\begin{align}\label{scaling_inter}
&g_m=g_0(-1)^me^{-im\pi c}\(\frac{u_0+1}{u_0-1}\)^{mc} \quad\quad\quad m>0 \\
&g_m=g_{-1}(-1)^{m+1}e^{i(m+1)\pi c}\(\frac{u_0+1}{u_0-1}\)^{(m+1)c}\quad m<-1.
\end{align}
The coefficients of the most singular part of (\ref{scaling_inter})
can be compared with the results from direct evaluation of
$G_{LO}(t=t_m(1+x),t')$. The procedure is analogous to that in the
previous section, with details included in Appendix \ref{apB}. Here we simply quote the results for $G_{LO}(t,x')$:
\begin{align}
&G_{LO}(t<t')\deq A_2e^{\frac{i\pi(d-1)m}{2}}\frac{1}{(-iy+\eps)^{c+1}},\quad n<0, \\
&G_{LO}(t>t')\deq A_2e^{\frac{-i\pi(d-1)(m+1)}{2}}\frac{1}{(iy+\eps)^{c+1}}, \quad n\ge 0,
\end{align}
with $A_2=2^c\frac{\Gm(c)}{\Gm(d)}t'{}^{-c}\(\frac{t_m}{u_0}\)^{\frac{d-1}{2}}\frac{\Gm(1+c)}{2\pi}i^c$.

The two results do not agree with each other at a first glance. The
disagreement can be traced back to the difference in the potential
used in two approaches. In the divergence matching procedure, we used
the potential $V_1=V_0\dlt(z-\frac{t}{u_0})$, while in 
 Appendix \ref{apB}, we use $V_2=V_0\dlt(\frac{t}{z}-u_0)$. The latter form of the potential is necessary in order to keep the scale invariance of theory.

To compare the two results on equal footing, we note that as $t''\to t_m$,
\ba\label{potentials}
V_2(t'',z'')=\frac{t_m}{u_0^2}V_1(t'',z'').
\ea
Using (\ref{potentials}) to convert one potential into the other, and using that 
$t_m=t'\frac{u_0}{u_0-1}\(\frac{u_0+1}{u_0-1}\)^m$, we can show that the two results on the most singular part of $G_{LO}$ along the mirror trajectory do agree with each other.

We may further compare the final results for the two-point correlator obtained
from the two different approaches. This time the difference in the
form of the potential does not cause a problem because the limit
$V_0\to\infty$ for both $V_1$ and $V_2$ gives the Dirichlet boundary
condition along the trajectory of the mirror. The two-point correlator
from the divergence matching procedure is obtained by inserting (\ref{scaling_inter}) to (\ref{correlator_f}), which gives:
\begin{align}\label{scaling_corr}
&\lag TO(t=\tb_m(1+x))O(t')\rag \no
&\deq -\frac{i\Gm(c)}{\sqrt{\pi}\Gm(c+\frac{1}{2})}\(\frac{1}{\tb_mt'}\)^c\frac{\Gm(c)}{\Gm(d)}\frac{\Gm(1+c)}{2\pi}e^{-i\pi(c-1)(m-1)}\sin\pi c\, B(2c,1-c)\frac{1}{(-x+i\eps)^{2c}}, \quad m\ge0,\\
&\lag TO(t=\tb_m(1+x))O(t')\rag \no
&\deq -\frac{i\Gm(c)}{\sqrt{\pi}\Gm(c+\frac{1}{2})}\(\frac{1}{\tb_mt'}\)^c\frac{\Gm(c)}{\Gm(d)}\frac{\Gm(1+c)}{2\pi}e^{i\pi(c-1)(m+1)}\sin\pi c\, B(2c,1-c)\frac{1}{(-x+i\eps)^{2c}}, \quad m\le-1.
\end{align}
The result of the two-point correlator from direct evaluation in Appendix \ref{apB} is:
\begin{align}
&\lag TO(t=\tb_m(1+x))O(t')\rag \no
&\deq(\tb_m t')^{-\frac{d+1}{2}}2^d(1-e^{i\pi d})\frac{\Gm(-d)\Gm(\frac{1+d}{2})}{\Gm(\frac{1-d}{2})\Gm(d)}e^{\frac{-i\pi(d-1)m}{2}}\frac{\Gm(1+d)}{e^{-i\pi(d+1)}(-x+i\eps)^{d+1}2i\pi}, &m\ge0,\\
&\lag TO(t=\tb_m(1+x))O(t')\rag \no
&\deq(\tb_m t')^{-\frac{d+1}{2}}2^d(1-e^{-i\pi d})\frac{\Gm(-d)\Gm(\frac{1+d}{2})}{\Gm(\frac{1-d}{2})\Gm(d)}e^{\frac{i\pi(d-1)m}{2}}\frac{\Gm(1+d)}{e^{-i\pi(d+1)}(x+i\eps)^{d+1}2i\pi}, &m\le-1.
\end{align}
We show in Appendix \ref{apB}  that the two approaches give exactly the same results!

\section{Non-local double trace deformation}

It has been established in \cite{rangamani} 
that a bulk Dirichlet boundary condition leads to a double trace deformation
on the field theory, based on earlier works \cite{double_tr,Kuperstein:2011fn}. The
double trace deformation can be non-local in the dual field theory,
generically breaking Lorentz invariance. We will give a concrete proposal for this deformation that matches
with our results for the singularity structure of the correlators.

Let us introduce in the field-theory Lagrangian a double trace deformation involving a scalar operator where the two insertions are separated by a relative constant time $T_0>0$:
\begin{equation}\label{eq:doubdef}
S_g=\int d^d x {\cal L}_0+g\int d^d x {\cal O}\left(t+T_0,\mathbf{x}\right){\cal O}\left(t,\mathbf{x}\right) ,
\end{equation}
where $g$ is coupling of the double trace.

In order to compute the time-ordered correlation function in the presence of the double trace deformation, we will perform a formal expansion in $g$. For economy, we will suppress the spatial dependence of operators and keep explicitly only the time dependence
\begin{equation}\label{eq:Tordcorr}
\left\langle T{\cal O}(t){\cal O}(0)\right\rangle_g =
\left\langle T{\cal O}(t){\cal O}(0)\right\rangle_0+g\int
d^d x'\, \left\langle T{\cal O}(t){\cal O}(0){\cal
      O}\left(t'+T_0\right){\cal
      O}\left(t'\right)\right\rangle_0+O(g^2) \, .
\end{equation}
Let us take advantage of the large-$N$ limit and treat ${\cal O}$ as an essentially free field. The logic behind this is that the correlator above is not just the connected component but it has contributions from the factorization in correlators with a smaller number of insertions. In the large-$N$ limit the connected four-point correlator is suppressed respect to these factorized contributions. Since the expectation value of the operator is zero, the leading contribution comes from the factorization in two-point functions. Let us assume for the moment $t>t'+T_0>t'>0$, then
\begin{equation}\label{eq:factorizcorr}
\left\langle T{\cal O}(t){\cal O}(0){\cal O}\left(t'+T_0\right){\cal O}\left(t'\right)\right\rangle_0\simeq \left\langle T{\cal O}(t){\cal O}\left(t'+T_0\right)]\right\rangle_0 \left\langle T{\cal O}\left(t'\right){\cal O}(0)\right\rangle_0+\cdots
\end{equation}
The dots involve a contraction of ${\cal O}(t)$ with ${\cal O}(0)$, so their contribution is just a one-loop renormalization (after integrating over $x'$) of the overall factor of the two-point function in the absence of the deformation. The term we have written explicitly is more interesting, it involves propagation from $0$ to $t'$ and from $t'+T_0$ to $t$. The integration over $t'$ in the full expression \eqref{eq:Tordcorr} suggests a convolution between propagators, however this is only possible if the two propagators connect at the same point. We can now take advantage of the Poincar\'e invariance of the undeformed theory, the propagator depends only on the difference in time, so we can shift the arguments $t'+T_0\to t'$, $t\to t-T_0$, so the first propagator in \eqref{eq:factorizcorr} now connects $t'$ to $t-T_0$. Now the two propagators connect at the same point at $t'$ and after integrating in $x'$, we just have the convolution of the two. As usual, the convolution of two propagators is a propagator connecting the external points, so we get a propagator between $0$ and $t-T_0$. We can repeat the argument for the remaining possibilities concerning time ordering, for instance if $0>t'+T_0>t'>t$, one propagator connects $t$ to $t'$ and the other $t'+T_0$ to $0$. In this case we can shift $t\to t+T_0$ and $t'\to t'+T_0$ in order to make the convolution. Therefore, 
\begin{align}\label{eq:Tordcorr2}
\notag\left\langle T{\cal O}(t){\cal O}(0)\right\rangle_g &= {\cal Z}_{\cal O}\left\langle T{\cal O}(t){\cal O}(0)\right\rangle_0+\\&g\int d^{d-1} { x}'\, \left\langle T{\cal O}(t-T_0){\cal O}(0)\right\rangle_0+g\int d^{d-1} {x}'\, \left\langle T{\cal O}(t+T_0){\cal O}(0)\right\rangle_0+O(g^2),
\end{align}
where we have introduced a renormalization factor ${\cal Z}_{\cal O}$ to account for the other contributions. 

The key point here is that if the two-point function of the undeformed theory has a singularity of the form
\begin{equation}\label{eq:Tordcorr3}
\left\langle T{\cal O}(t){\cal O}(0)\right\rangle_0\sim \frac{1}{t^{2\Delta}},
\end{equation}
then, in the presence of the double trace deformation, there are new singularities that are displaced in time by $\pm T_0$
\begin{equation}\label{eq:Tordcorr4}
\left\langle T{\cal O}(t){\cal O}(0)\right\rangle_g\sim \frac{1+O(g)}{t^{2\Delta}}+\frac{g+O(g^2)}{(t-T_0)^{2\Delta}}+\frac{g+O(g^2)}{(t+T_0)^{2\Delta}}+O(g^2).
\end{equation}
We can follow the same procedure to see the effect of $O(g^2)$ terms, they introduce singularities that are displaced in time by $\pm 2 T_0$. In general, at order $O(g^n)$, new singularities displaced by $\pm n T_0$ appear. Taking into account the full infinite sum, we find the same singularity structure as for the AdS/CFT setup with a static mirror sitting at $z=T_0$. This suggests that the non-local double trace deformation in \eqref{eq:doubdef} is the right one to describe Dirichlet boundary conditions in AdS. The straightforward generalization to a moving mirror would be to make $T_0$ a time-dependent function, given precisely by the trajectory of the mirror. From the field theory point of view this is a much more complicated problem, using AdS/CFT we can find a simple answer just applying geometric optics.

The analysis we have done is incomplete, since we have studied only two-point functions. In the presence of a mirror, higher point functions in the field theory side might be modified in a way that the double trace deformation does not capture. If that is the case, presumably higher order multi-trace deformations would be needed, the methods presented here may also be useful to determine them. 

\section{Spectral decomposition and scaling properties}\label{sec}

To understand the physical implication of the change of amplitude we found in the case of the scaling mirror, it is useful to look at the spectrum of the field theory with non-local double-trace deformation. We start with the simple case of static mirror. The spectrum is given by the normalizable modes. In the WKB approximation, which contains the UV mode, is found in Appendix \ref{apC} to be discrete equal distant resonances: $\omg=\pm\omg_n\equiv\pm\frac{n\pi}{z_s}$. The corresponding spectral density is given by
\begin{align}
\rho(\mu)\sim\sum_n\omg_n^d\(\dlt(\mu-\omg_n)-\dlt(\mu+\omg_n)\),
\end{align}
the power is fixed by dimension.
A general two point correlator has the following representation:
\begin{align}\label{spectral}
\left\langle {\cal O}(t){\cal O}(t')\right\rangle=\int_{-\infty}^{+\infty}d\omg d\mu\rho(\mu)\frac{e^{-i\omg(t-t')}}{\omg+\mu},
\end{align}
The retarded and advanced correlators are obtained shifting the frequency by $\pm i\epsilon$, while the Wightman correlators are obtained by shifting the time. The time ordered correlator is a combination of those. Taking into account that the spectrum is discrete and applying the residue theorem in evaluating the integral, we obtain: 
\begin{align}\label{normalizable}
\left\langle {\cal O}(t){\cal O}(t')\right\rangle=\sum_n\phi_n(t)\phi_n^*(t'),
\end{align}
with $\phi_n(t)\sim e^{-i\omg_nt}\omg_n^{\frac{d}{2}}$.
With an equidistant spectrum, it is easy to check that $\phi_n(t-2z_s)=\phi_n(t)$, which leads to the periodic behavior of the two-point correlator,
\begin{align}
\left\langle {\cal O}(t-2mz_s){\cal O}(t')\right\rangle=\left\langle {\cal O}(t){\cal O}(t)\right\rangle.
\end{align}
Therefore in the result of $\left\langle {\cal O}(t){\cal O}(t')\right\rangle$, a singularity of the form $\frac{r}{(t-t')^{2c}}$ is always accompanied by singularities of the form $\frac{r}{(t-2mz_s-t')^{2c}}$, corresponding a periodic behavior of the returning signal.

Now we consider a falling mirror with scaling trajectory, we can still use \eqref{normalizable} as a good representation of the correlator, but now $\phi_n(t)=t^{\lam_n-c}\lam_n^{\frac{d}{2}}$. $\lam_n=\frac{2i\pi n}{\ln\frac{u_0+1}{u_0-1}}$ is obtained from the appendix of \cite{ELN}. Similar to the case of a static mirror, we find:
\begin{align}
\phi_n(t/\frac{u_0+1}{u_0-1})=\(\frac{u_0+1}{u_0-1}\)^c\phi_n(t),
\end{align}
which leads to the identity for the correlator
\begin{align}\label{scalerelation}
\left\langle {\cal O}(t/\frac{u_0+1}{u_0-1}){\cal O}(t')\right\rangle=\(\frac{u_0+1}{u_0-1}\)^c\left\langle {\cal O}(t){\cal O}(t')\right\rangle.
\end{align}
If we expand the correlators appearing on both sides of \eqref{scalerelation} around the singularities:
\begin{equation}
\left\langle {\cal O}(t){\cal O}(t')\right\rangle=\cdots+ \frac{r_0}{\left(t-t' \right)^{2 c}}+\frac{r_1}{\left(t-\frac{u_0+1}{u_0-1}t' \right)^{2 c}}+\cdots,
\end{equation}
and
\begin{equation}
\left\langle {\cal O}(t/\frac{u_0+1}{u_0-1}){\cal O}(t')\right\rangle= \cdots+ \frac{r_0}{\left(t/\frac{u_0+1}{u_0-1}-t' \right)^{2 c}}+\frac{r_1}{\left(t/\frac{u_0+1}{u_0-1}-\frac{u_0+1}{u_0-1}t' \right)^{2 c}}+\cdots.
\end{equation}
Then,
\begin{equation}
\left\langle {\cal O}(t/\frac{u_0+1}{u_0-1}){\cal O}(t')\right\rangle=\left(\frac{u_0+1}{u_0-1} \right)^{2 c}\left[ \cdots+ \frac{r_0}{\left(t-\frac{u_0+1}{u_0-1}t' \right)^{2 c}}+\frac{r_1}{\left(t-\left(\frac{u_0+1}{u_0-1}\right)^2 t' \right)^{2 c}}+\cdots\right]
\end{equation}
Comparing with \eqref{scalerelation}, the equality is true if
\begin{equation}
r_1=\left(\frac{u_0+1}{u_0-1} \right)^{c}r_0
\end{equation}
or in general
\begin{equation}
r_{m+1}=\left(\frac{u_0+1}{u_0-1} \right)^{c}r_m
\end{equation}
which agrees with our previous analysis.

\section{Conclusion and Outlook}

We have established a general procedure for computing the singularity
structure of spatially integrated unequal-time correlators.  The
procedure is applicable to a bulk scalar in AdS space subject to a
Dirichlet boundary condition along arbitrary time-like trajectories.  The Dirichlet boundary condition is interpreted as non-local double trace deformation to the dual field theory, which generically leads to the emergence of new singularities.

An immediate application of the current procedure is to the
gravitational collapse model of thermalization. Although the current
procedure is formulated in a pure AdS background, it should be readily
extendable to the case of thermal AdS \footnote{Some complications due
  to using thermal AdS propagators are possible, however the
  calculations will be simplified by using spatially integrated
  propagators.}. This can then be used to study the unequal-time
two-point correlator in the far-from-equilibrium regime in the
gravitational collapse model. 
Similar patterns of singularities are expected in the resulting
two-point correlator. The separation and magnitude of the
singularities contains valuable information on the temporal decoherence
of the gauge fields in the thermalization process. These results will be
complementary to the study of equal-time correlators as for instance
considered in \cite{11authors},  and will provide concrete information
useful for understanding the physics of thermalization in strongly coupled gauge theory.

Recently, two interesting phenomena have been found in the context of
global AdS, which is dual to CFT on a sphere. The first one is the
existence of undamped oscillation modes in a thermal state
\cite{mcgreevy}. It implies that certain modes will never actually
thermalize in a CFT on a sphere. It was argued that the existence of
the oscillation modes originates from the evenly spaced spectrum of
the corresponding operator. Our two examples have shown similar
behavior: the modes for the static mirror ($\omg_n$) and for the
scaling mirror ($\lam_n$) are evenly spaced, and we have found that
singularities at arbitrary late time appear periodically in the first
case and quasi-periodically in the second case. It would be
interesting to see whether this persists in a realistic model of
thermalization. The second phenomenon is the non-perturbative
instability of global AdS due to the evenly spaced spectrum
\cite{horowitz}. While our model is formulated in the Poincar\'{e}
patch, the mirror provides another boundary, which effectively
produces a confining box in the Poincar\'{e} patch. The
non-perturbative instability of global AdS is due to the resonant
coupling between modes of different frequencies, which is possible
because they have simple ratios. In this respect, the model we
consider is similar in that respect to global AdS, so in principle the
same kind of instability could be present 
in the case of a static mirror. It will  be interesting to explore this possibility.

Another observation for the moving mirror is that the form of the correlator in the scaling
mirror is similar to the correlators found in ageing systems (for a
review see \cite{cond-mat/0703466}) that describe some
out-of-equilibrium states with slow dynamics in condensed matter, like
for instance a glass. In those systems, two-point functions do not
only depend on the time difference $t-t'$ between the two insertion points,
but also have a power-like dependence on $t/t'$. For this reason the number of scaling exponents necessary
to fully determine the system is larger than in the usual equilibrium
states. In the holographic model presented here, we also observe a 
power-like dependence on the initial time, and the exponents for $t-t'$ and
$t/t'$ are fixed by the dimension of the dual operator, at least in
the massless case. It would interesting to explore how far this
connection between ageing systems and holography can reach.

Finally, it is worth stressing that the results obtained in this work rely on
the large $N_c$ limit. The relaxation of this limit might be important
for thermalization in quantum field theory, as described for instance
in  \cite{berges,HLR}. It is an interesting challenge to see how the finite $N_c$ effect may change the results in this work.

\section*{Acknowledgements}

S.~L.~is grateful to the Alexander von Humboldt Foundation for a
fellowship. C.H.~would like to thank Bom Soo Kim for interesting discussions. 
This work has been supported in part by the `Excellence
Cluster for Fundamental Physics: Origin and Structure of the
Universe'. C.H.~was supported in part by the 
Israel Science Foundation (grant number 1468/06).

\appendix

\section{Integrals involving $i\eps$}\label{apA}

\ba\label{eq6}
&&\int_{-\infty}^\infty dy\frac{1}{(-iy+\eps)^{c+1}}\ln(y-x-i\eps')=\frac{-2i\pi}{i^{c+1}}\frac{1}{(-x-i\eps'')^cc} \no
&&\int_{-\infty}^\infty dy\frac{1}{(iy+\eps)^{c+1}}\ln(x-y-i\eps')=\frac{-2i\pi}{i^{c+1}}\frac{1}{(x-i\eps'')^cc} \no
&&\int_{-\infty}^\infty dy\frac{1}{(-iy+\eps)^{c+1}}\ln(x-y-i\eps')=\int_{-\infty}^\infty dy\frac{1}{(iy+\eps)^{c+1}}\ln(y-x-i\eps')=0 \no
&&\int_{-\infty}^\infty dy\frac{1}{(-iy+\eps)^{c+1}}\ln(-(x-y)^2+i\eps')=\frac{-2i\pi}{i^{c+1}}\frac{1}{(-x-i\eps'')^cc} \no
&&\int_{-\infty}^\infty dy\frac{1}{(iy+\eps)^{c+1}}\ln(-(x-y)^2+i\eps')=\frac{-2i\pi}{i^{c+1}}\frac{1}{(x-i\eps'')^cc}.
\ea
\ba\label{eq_int}
&&\int_{-\infty}^\infty\frac{1}{(-iy+\eps)^{c+1}}\frac{1}{(x-y+i\eps)^c}=(e^{-i\pi c}-e^{i\pi c})i^{c+1}\frac{1}{(x+i\eps)^{2c}}B(2c,1-c) \no
&&\int_{-\infty}^\infty\frac{1}{(iy+\eps)^{c+1}}\frac{1}{(y-x+i\eps)^c}=(e^{-i\pi c}-e^{i\pi c})i^{c+1}\frac{1}{(-x+i\eps)^{2c}}B(2c,1-c) \no
&&\int_{-\infty}^\infty\frac{1}{(-iy+\eps)^{c+1}}\frac{1}{(y-x+i\eps)^c}=0 \no
&&\int_{-\infty}^\infty\frac{1}{(iy+\eps)^{c+1}}\frac{1}{(x-y+i\eps)^c}=0 .
\ea

A few comments about these integrals are helpful: i) we have extended the integration
range of $y$ to infinity. This is justified as far as the singular part in $x$ is 
concerned and the integral is convergent. It turns out the results of the integrals
in (\ref{eq6}) and (\ref{eq_int}) contain no regular part. If it were not the case, we should have
discarded the regular parts; ii) The vanishing of the last two integrals is 
due to the fact that the branch cuts associated with the integrand can be chosen
to lie at a single side of the half plane (upper or lower), thus the integrals are 
guaranteed to vanish by a proper choice of contour; iii) $\eps''$ is a combination
of $\eps$ and $\eps'$, but the precise relation is not important, as we are interested
in the limit $\eps\to 0$. We will simply omit the primes.

\section{Explicit evaluations of the correlators: static mirror}\label{apC}

In order to compute the correlator, we write a general scalar field in the bulk in frequency space. Since we focus on the spatially integrated correlator, the scalar field depends on $t$ and $z$ only.
\begin{align}
\phi(t,z)=\int g(\omg)\pht(\omg,z)e^{-i\omg\Dlt t}d\omg,
\end{align}
where $\Dlt t=t-t'$. $g(\omg)$ is the arbitrary Fourier coefficient and $\pht(\omg,z)$ satisfies the Laplacian equation in the bulk in the presence of a potential $V=V_0\dlt(z-z_s)$. $\pht(\omg,z)$ is easily solved by:
\begin{align}
&\pht(\omg,z)=z^{\frac{d}{2}}\(A J_{\frac{d}{2}}(\lv\omg\rv z)+B J_{-\frac{d}{2}}(\lv\omg\rv z)\), &z<z_s\\
&\pht(\omg,z)=z^{\frac{d}{2}}CH_{\frac{d}{2}}^{(1)}(|\omg|z) &z>z_s.
\end{align}
The solutions above and below the potential are matched in a standard way:
\begin{align}\label{static_matching}
\left\{\begin{array}{l@{\quad\quad}}
AJ_{\frac{d}{2}}(|\omg|z_s)+BJ_{-\frac{d}{2}}(|\omg|z_s)=CH_{\frac{d}{2}}^{(1)}(|\omg|z_s)\\
z_s^2|\omg|\bigg[AJ_{\frac{d}{2}-1}(|\omg|z_s)-BJ_{-\frac{d}{2}+1}(|\omg|z_s)-CH_{\frac{d}{2}-1}^{(1)}(|\omg|z_s)\bigg]=V_0CH_{\frac{d}{2}}^{(1)}(|\omg|z_s)
\end{array}
\right.
\end{align}
It is not difficult to convince ourselves that $C$ is $\frac{1}{V_0}$ suppressed compared to $A,B$ and to the LO in $V_0$, we have
\ba
AJ_{\frac{d}{2}}(|\omg|z_s)+BJ_{-\frac{d}{2}}(|\omg|z_s)=0.
\ea
We may take $A=1$ and $B=\ka(|\omg|)=-\frac{J_{\frac{d}{2}}(\lv\omg\rv z_s)}{J_{\frac{d}{2}}(\lv\omg\rv z_s)}$ for specificness. Using the second line of (\ref{static_matching}), we obtain an explicit expression for $C$ at order $\frac{1}{V_0}$:
\begin{align}
C=\frac{z_s^2|\omg|}{V_0}\frac{J_{\frac{d}{2}-1}(|\omg|z_s)-\ka(|\omg|)J_{-\frac{d}{2}+1}(|\omg|z_s)}{H_{\frac{d}{2}}^{(1)}(|\omg|z_s)}.
\end{align}
The scalar wave on the mirror is simply given by:
\begin{align}
\pht(\omg,z=z_s)&=Cz^{\frac{d}{2}}H_{\frac{d}{2}}^{(1)}(|\omg|z_s) \no
&=\frac{z_s^{c+1}}{V_0}\sqrt{\frac{2|\omg|}{\pi}}\frac{\cos\frac{\pi(d-1)}{2}}{\cos\(|\omg|z_s+\frac{\pi(d-1)}{4}\)},
\end{align}
where in the last step, we have taken $|\omg|\to\infty$ to simplify the expression. This will not affect our results on the most singular part of $G_{LO}$, which is supposed to come from the UV physics.
To calculate $G_{LO}$, we notice that $\phi(t,z)$ has the following expansion:
\begin{align}
\phi(t,z)=\phi_0(t)+\cdots\phi_d(t)z^d+\cdots,
\end{align}
where the coefficients $\phi_0$ and $\phi_d$ can be expressed by
\begin{align}\label{omg_rep}
&\phi_0(\Dlt t)=\int g(\omg)\ka(\omg)\frac{1}{\Gm(1-\frac{d}{2})}\(\frac{2}{|\omg|}\)^{d/2}e^{-i\omg\Dlt t}d\omg \\
&\phi_d(\Dlt t)=\int g(\omg)\frac{1}{\Gm(1+\frac{d}{2})}\(\frac{|\omg|}{2}\)^{d/2}e^{-i\omg\Dlt t}d\omg.
\end{align}
On the other hand,
\begin{align}\label{omg_rep_Glo}
\phi(\Dlt t,z=z_s)=\int g(\omg)Cz^{\frac{d}{2}}H_{\frac{d}{2}}^{(1)}(|\omg|z_s)e^{-i\omg\Dlt t}d\omg .
\end{align}
An easy way to calculate $G_{LO}$ is to set $\phi_0(\Dlt t)=\dlt(\Dlt t)$, which allows us to solve for $g(\omg)$. Plugging $g(\omg)$ to (\ref{omg_rep_Glo}), we obtain $G_{LO}(\Dlt t)$, in the following frequency space:
\begin{align}\label{Glo_static}
G_{LO}(\Dlt t)=\int\frac{1}{2\pi}\frac{\sqrt{\frac{2}{\pi}}\Gm(1-\frac{d}{2})}{2^{\frac{d}{2}}}\frac{|\omg|^cz_s^{c+1}\cos\frac{\pi(d-1)}{2}}{\cos\(|\omg|z_s-\frac{\pi(d+1)}{4}\)}e^{-i\omg\Dlt t}d\omg
\end{align}

We use the residue theorem to evaluate (\ref{Glo_static})\footnote{The residue theorem is not directly applicable in the presence of $|\omg|$. A proper way to treat it is to make the substitution $|\omg|\to\sqrt{\omg^2+\eta^2}$ and consider the limit $\eta\to 0$. The square root brings in branch cuts, but one can show the contribution from the contour wrapping the branch cuts vanishes. As a net result, we may focus on the isolated poles only}. The poles of the
integrand is given by the zeros of $\cos(|\omg|z_s-\frac{\pi(d+1)}{4})$, which
are located at $\omg=\pm\omg_n$, with $\omg_n=\frac{\(n+\frac{d+3}{4}\)\pi}{z_s}\,(n=0,1,\cdots)$. The poles lie along the real axis symmetrically. We need to
deform the integration contour to avoid the poles. The time-ordered correlator
can be obtained with the contour shifted slightly counter-clockwise. This corresponds to the following substitution:
\begin{align}
\omg\to\omg(1+i\eps),\,t\to t(1-i\eps).
\end{align}
For $\Dlt t<0$, we close the integration contour upwards. The poles on the 
negative real axis contribute. The residue come from $\frac{1}{\cos(|\omg|z_s-\frac{\pi(d+1)}{4})}$ are given by:
\ba
\text{res}\(\frac{1}{\cos(|\omg|z_s-\frac{\pi(d+1)}{4})}\)=\frac{(-1)^n}{z_s}.
\ea
Summing over the residues, we obtain
\begin{align}
&G_{LO}(\Dlt t)=\frac{i\pi^c\sqrt{\frac{2}{\pi}}\Gm(1-\frac{d}{2})}{2^{\frac{d}{2}}}\cos\frac{\pi(d-1)}{2}e^{\frac{i\pi(d+3)\Dlt t}{4z_s}}(n+\frac{d+3}{4})^ce^{in\pi(\Dlt t/z_s-1)-\eps} \no
&=\frac{i\pi^c\sqrt{\frac{2}{\pi}}\Gm(1-\frac{d}{2})}{2^{\frac{d}{2}}}\cos\frac{\pi(d-1)}{2}e^{\frac{i\pi(d+3)\Dlt t}{4z_s}}\Phi(w,s,v),
\end{align}
where $w=e^{i\pi(\Dlt t/z_s-1)-\eps}$, $s=-c$ and $v=\frac{d+3}{4}$. $\Phi(w,s,v)$ is the Lerch transcendent function. the most singular part of $G_{LO}$ follows from an expansion of the Lerch function:
\begin{align}
\Phi(w,s,v)\deq\Gm(1+c)e^{-\frac{i\pi(\Dlt t-(2m+1)z_s)}{z_s}\frac{d+3}{4}}\(-i\pi\frac{\Dlt t-(2m+1)z_s}{z_s}\)^{-c-1}.
\end{align}
In the limit $t=t_m(1+x),\;x\to 0$, we obtain the most singular part of $G_{LO}$ as
\begin{align}\label{Glo_tg0}
G_{LO}(\Dlt t)=\frac{i\sqrt{\frac{2}{\pi}}\Gm(1-\frac{d}{2})}{2^{\frac{d}{2}}\pi}\cos\frac{\pi(d-1)}{2}e^{\frac{i\pi(1+2m)}{2}}\frac{1}{(-ix+\eps)^{c+1}}\(\frac{z_s}{t_m}\)^{c+1}.
\end{align}
For $\Dlt t>0$, we close the contour downwards and include contributions from poles on the positive real axis. We obtain the following expression for $G_{LO}$, after summing over the residues:
\begin{align}
&G_{LO}(\Dlt t)=\frac{i\pi^c\sqrt{\frac{2}{\pi}}\Gm(1-\frac{d}{2})}{2^{\frac{d}{2}}}\cos\frac{\pi(d-1)}{2}e^{-\frac{i\pi(d+3)\Dlt t}{4z_s}}(n+\frac{d+3}{4})^ce^{-in\pi(\Dlt t/z_s-1)-\eps} \no
&=\frac{i\pi^c\sqrt{\frac{2}{\pi}}\Gm(1-\frac{d}{2})}{2^{\frac{d}{2}}}\cos\frac{\pi(d-1)}{2}e^{-\frac{i\pi(d+3)\Dlt t}{4z_s}}\Phi(w,s,v),
\end{align}
with $w=e^{-i\pi(\Dlt t/z_s-1)-\eps}$, $s=-c$ and $v=\frac{d+3}{4}$.
The expansion of the Lerch function gives the following most singular part of $G_{LO}$:
\begin{align}\label{Glo_tl0}
G_{LO}(\Dlt t)=\frac{i\sqrt{\frac{2}{\pi}}\Gm(1-\frac{d}{2})}{2^{\frac{d}{2}}\pi}\cos\frac{\pi(d-1)}{2}e^{-\frac{i\pi(1+2m)}{2}}\frac{1}{(ix+\eps)^{c+1}}\(\frac{z_s}{t_m}\)^{c+1}.
\end{align}
We can compare \eqref{Glo_tg0} and \eqref{Glo_tl0} with the results obtained from the recursion equations. The results are in perfect agreement.

We can further compute the two-point correlator and compare the results with the output from recursion equations. The evaluation of the two-point correlator closely resembles the evaluation of $G_{LO}$. It simply amounts to the evaluation of the following integral:
\begin{align}\label{G_static}
G(\Dlt t)=\int\frac{1}{2\pi}\frac{1}{\ka(\omg)}\frac{\Gm(1-\frac{d}{2})}{\Gm(1+\frac{d}{2})}\(\frac{|\omg|}{2}\)^de^{-i\omg\Dlt t}d\omg.
\end{align}
We will not repeat the steps in doing the integral, but just state the final result. For $\Dlt t<0$, we have
\begin{align}
G(\Dlt t<0)=-\frac{i\cos\frac{d+1}{2}\pi}{z_s}\frac{\Gm(1-\frac{d}{2})}{\Gm(1+\frac{d}{2})}\(\frac{\pi}{2z_s}\)^de^{\frac{i\frac{d+3}{4}\pi\Dlt t(1-i\eps)}{z_s}}\Phi(w,s,v),
\end{align}
where $w=e^{\frac{i\pi\Dlt t(1-i\eps)}{z_s}}=e^{\frac{i\pi\Dlt t}{z_s}-\eps}$, $v=\frac{d+3}{4}$, $s=-d$. 
where we have used the same definition as before $x$: $t=\tb_m(1+x)$ and $\tb_m=t'+2mz_s$.
The most singular part of the correlator is given by
\begin{align}\label{Gtg0}
G(\Dlt t<0)\deq -i\cos\frac{d+1}{2}\pi\frac{\Gm(1-\frac{d}{2})}{\Gm(1+\frac{d}{2})}\(\frac{\pi}{2}\)^d\Gm(1+d)e^{im\pi\frac{d+3}{2}}\frac{1}{(e^{-i\pi/2}\pi\tb_m)^{2c}}\frac{1}{(x+i\eps)^{2c}}.
\end{align}
For $\Dlt t>0$, we have
\begin{align}
G(\Dlt t<0)=-\frac{i\cos\frac{d+1}{2}\pi}{z_s}\frac{\Gm(1-\frac{d}{2})}{\Gm(1+\frac{d}{2})}\(\frac{\pi}{2z_s}\)^de^{-\frac{i\frac{d+3}{4}\pi\Dlt t(1-i\eps)}{z_s}}\Phi(w,s,v),
\end{align}
where $w=e^{-\frac{i\pi\Dlt t(1-i\eps)}{z_s}}=e^{-\frac{i\pi\Dlt t}{z_s}-\eps}$, $v=\frac{d+3}{4}$, $s=-d$.
The most singular part of the correlator is given by
\begin{align}\label{Gtl0}
G(\Dlt t>0)\deq -i\cos\frac{d+1}{2}\pi\frac{\Gm(1-\frac{d}{2})}{\Gm(1+\frac{d}{2})}\(\frac{\pi}{2}\)^d\Gm(1+d)e^{-im\pi\frac{d+3}{2}}\frac{1}{(e^{-i\pi/2}\pi\tb_m)^{2c}}\frac{1}{(-x+i\eps)^{2c}}.
\end{align}
We can verify that (\ref{Gtg0}), (\ref{Gtl0}) and (\ref{correlator_s}) again agree with each other!

\section{Explicit evaluations of the correlators: scaling trajectory}\label{apB}

The scaling trajectory corresponds to the potential $V=V_0\dlt(\frac{t}{z}-u_0)$. As familiar from quantum mechanics, the solution to the wave equation in the bulk is solved by a matching of the wave
above and below the mirror. We closely follow the notation of \cite{ELN}. 
For the eigenmode $\phi(u,v)=v^{\frac{\lam}{2}}f(u)$, the
incoming/outgoing solutions are given by:
\ba
f_{\pm}(u)=u^{-\frac{\lam}{2}-\frac{d-1}{4}}(u\mp 1)^\lam F(\frac{1-d}{2},\frac{1+d}{2};\pm\lam+1,\frac{1-u}{2}).
\ea
Above the mirror, the solution is a combination of incoming and outgoing waves 
$f(u)=Af_+(u)+Bf_-(u)$ and below the mirror there is only incoming component
$f(u)=Cf_+(u)$\footnote{The time-ordered correlator actually requires incoming wave for $\frac{\lam}{i}>0$ and outgoing wave for $\frac{\lam}{i}<0$. However, it is a short exercise to show that either incoming or outgoing wave below the mirror gives rise to the same $\phi_\lam(t,z=t/u_0)$, thus the following computation is not affected.}. Matching on the mirror gives:
\ba
\left\{\begin{array}{l}
Af_+'(u_0)+Bf_-'(u_0)=Cf_+'(u_0)+\frac{V_0}{u_0^2-1}Cf_+(u_0) \\
Af_+(u_0)+Bf_-(u_0)=Cf_+(u_0)
\end{array}
\right.
\ea

We choose the normalization such that $A,\,B\sim 1$, $C\sim\frac{1}{V_0}$ as
$V_0\rightarrow\infty$. We work to the LO in $V_0$ from now on.  We choose
\ba
&&A=(u_0+1)^\lam F(\frac{1-d}{2},\frac{1+d}{2};-\lam+1,\frac{1-u_0}{2}) \no
&&B=-(u_0-1)^\lam F(\frac{1-d}{2},\frac{1+d}{2};\lam+1,\frac{1-u_0}{2}) \nonumber,
\ea
according to $Af_+(u_0)+Bf_-(u_0)=0$. It follows as $\lam\rightarrow\infty$,
\ba
C=\frac{u_0^2-1}{V_0}\frac{Af_+'(u_0)+Bf_-'(u_0)}{f_+(u_0)}\rightarrow \frac{2\lam(u_0+1)^\lam}{V_0}.
\ea

\begin{figure}
\includegraphics[width=0.5\textwidth]{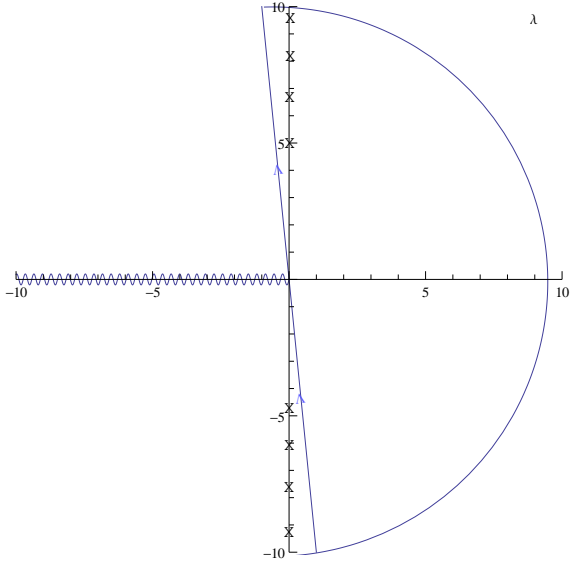}
\caption{\label{mellin}(color online) The isolated singularities lie symetrically along the imaginary axis. A branch cut extends from the origin to negative infinity. The integration contour is shifted slightly counter-clockwise for the time-ordered corrletor.}
\end{figure}

The bulk-boundary propagator is built as follows:
\ba\label{Gscaling}
&&\dlt(t-t')=\phi^0(t)=\int K(\lam,A,B)t^{\lam+\frac{d-1}{2}}g(\lam)d\lam \no
&&G(t,z,t')=\int C\phi_\lam(t,z)g(\lam)d\lam \no
\Rightarrow &&G(t,z,t')=\frac{1}{2\pi i}\int d\lam\frac{C\phi_\lam(t,z)}{K(\lam,A,B)}t'{}^{-\lam-\frac{d+1}{2}}.
\ea
We first compute $G(t,z,t')$ along the mirror trajectory, that is we set $z=\frac{t}{u_0}$. As $\lam\rightarrow\infty$, (\ref{Gscaling}) leads to the following representation of $G_{LO}$:
\begin{align}\label{eq1}
G_{LO}(t,t')\rightarrow \frac{A_1}{2\pi i}\int\frac{\lam(u_0+1)^\lam\left(1-\frac{1}{u_0}\right)^\lam}{\frac{\Gm(\lam+1)}{\Gm(\lam+\frac{d+1}{2})}(u_0+1)^\lam-\frac{\Gm(-\lam+1)}{\Gm(-\lam+\frac{d+1}{2})}(u_0-1)^\lam}\(\frac{t}{t'}\)^\lam t'{}^{-\frac{d+1}{2}}\(\frac{t}{u_0}\)^\frac{d-1}{2}d\lam,
\end{align}
with $A_1=2^\frac{d+1}{2}\frac{\Gm(\frac{d+1}{2})}{\Gm(d)}$. The integral in (\ref{eq1})
is not well defined. Certain $i\eps$ prescription is needed. Since we are interested in
the time-ordered correlator, we deform the integration contour as follows: 
$\lam\rightarrow \lam(1+i\eps)$, together with $\ln\frac{t}{t'}\rightarrow\ln\frac{t}{t'}(1-i\eps)$, to ensure the reality of the Fourier factor $\(\frac{t}{t'}\)^\lam$. The integration
contour is shown in Fig.\ref{mellin}.

For $t<t'$, we can close the contour on the right half plane to include residues on
the positive imaginary axis. Similar to the appendix of \cite{ELN}, we have:
\ba
G_{LO}(t<t')=\frac{A_1}{a}t'{}^{-\frac{d+1}{2}}\(\frac{t}{u_0}\)^{\frac{d-1}{2}}
e^{\frac{i\pi b(d-1)}{2a}}\(\frac{2i\pi}{a}\)^c\Phi(e^{\frac{2i\pi b}{a}},-c,\frac{d-1}{4}),
\ea
with $a=\ln\frac{u_0+1}{u_0-1},\,b=\ln\frac{t}{t'}+\ln\frac{u_0-1}{u_0},\,c=\frac{d+1}{2}$.
$\Phi$ is the Lerch transcendent function.
The $i\eps$ prescription above gives $b\rightarrow b+i\eps$. It is
important for fixing the ambiguity in the argument of the Lerch transcendent function.
Using the following decomposition formula for Lerch transcendent:
\ba\label{lerch}
\Phi(w,s,v)=\Gm(1-s)w^{-v}\(\ln\frac{1}{w}\)^{s-1}+w^{-v}\sum_{r=0}^\infty \zeta(s-r,v)\frac{\(\ln w\)^r}{r!}
\ea
Define $t_n=t'\frac{u_0}{u_0-1}\(\frac{u_0+1}{u_0-1}\)^n$. It is easy to see $t_n$ are
just the points where the bouncing light ray hit the mirror. As $t\to t_n$, $\ln w\to 0$,
the first term in (\ref{lerch}) is singular while the sum is regular.
We obtain the singular part of $G_{LO}(t<t')$ given by:
\ba
G_{LO}(t<t')\deq \frac{A_1}{a^{c+1}}t'{}^{-\frac{d+1}{2}}\(\frac{t}{u_0}\)^{\frac{d-1}{2}} e^{\frac{i\pi (d-1)n}{2}}\frac{\Gm(1+c)}{2\pi}\frac{i^c}{(-if+\eps)^{c+1}},
\ea
where $f=\frac{b}{a}-n=\frac{\ln\frac{t}{t_n}}{a},\,|f|<\frac{1}{2},\,n<0$. The 
symbol implies that the equality holds as far as the singular part is concerned.
In writing $i^c$, we
have chosen $\arg(\pm i)=\pm\frac{\pi}{2}$ to simplify the notation.

For $t>t'$, we make a change of variable $\lam\to-\lam$ in (\ref{eq1}). The latter now
becomes:
\ba\label{eq2}
G_{LO}(t>t')\rightarrow \frac{A_1}{2\pi i}\int\frac{-\lam u_0^\lam}{\frac{\Gm(-\lam+1)}{\Gm(-\lam+\frac{d+1}{2})}(u_0-1)^\lam-\frac{\Gm(\lam+1)}{\Gm(\lam+\frac{d+1}{2})}(u_0+1)^\lam}\(\frac{t'}{t}\)^\lam t'{}^{-\frac{d+1}{2}}\(\frac{t}{u_0}\)^\frac{d-1}{2}d\lam.
\ea
We can again close the contour on the right half plane to obtain:
\ba
G_{LO}(t>t')\deq \frac{A_1}{a^{c+1}}t'{}^{-\frac{d+1}{2}}\(\frac{t}{u_0}\)^{\frac{d-1}{2}} e^{-\frac{i\pi (d-1)(n+1)}{2}}\frac{\Gm(1+c)}{2\pi}\frac{i^c}{(if+\eps)^{c+1}},
\ea
with $f=\frac{\ln\frac{t}{t_n}}{a},\,|f|<\frac{1}{2},\,n\ge 0$.

Define y as $t=t_n(1+y)$. The dimensionless ratio $y$ measured the closeness of 
$t$ to $t_n$. To the LO in $y$, $\ln\frac{t}{t_n}\deq y$, thus  we obtain
the most singular part of $G(t,t')$ as follows:
\ba\label{Gtr_LO}
\label{eq3}
G_{LO}(t<t')\deq A_2e^{\frac{i\pi(d-1)n}{2}}\frac{1}{(-iy+\eps)^{c+1}}\quad n<0 \\
\label{eq4}
G_{LO}(t>t')\deq A_2e^{\frac{-i\pi(d-1)(n+1)}{2}}\frac{1}{(iy+\eps)^{c+1}} \quad n\ge 0,
\ea
with $A_2=A_1t'{}^{-\frac{d+1}{2}}\(\frac{t_n}{u_0}\)^{\frac{d-1}{2}}\frac{\Gm(1+c)}{2\pi}i^c$.

We also want to compute the two-point correlator by direct evaluation of the 
$\lam$-integral in Mellin representation. It is not difficult to see that the time-ordered correlator for $t>t'$ and $t<t'$ are basically the contribution corresponding to the lower sign and upper sign respectively in the retarded correlator in (3.19) of \cite{ELN}. The most singualr parts are given by the following:
\begin{align}\label{eq12}
&t>t':\; T\lag O(t=\tb_m(1+x))O(t')\rag \no
&\deq(\tb_m t')^{-\frac{d+1}{2}}2^d(1-e^{i\pi d})\frac{\Gm(-d)\Gm(\frac{1+d}{2})}{\Gm(\frac{1-d}{2})\Gm(d)}e^{\frac{-i\pi(d-1)m}{2}}\frac{\Gm(1+d)}{e^{-i\pi(d+1)}(-x+i\eps)^{d+1}2i\pi} \\
&t<t':\; T\lag O(t=\tb_m(1+x))O(t')\rag \no
&\deq(\tb_m t')^{-\frac{d+1}{2}}2^d(1-e^{-i\pi d})\frac{\Gm(-d)\Gm(\frac{1+d}{2})}{\Gm(\frac{1-d}{2})\Gm(d)}e^{\frac{i\pi(d-1)m}{2}}\frac{\Gm(1+d)}{e^{-i\pi(d+1)}(x+i\eps)^{d+1}2i\pi}.
\end{align}





\end{document}